\documentclass[iop]{emulateapj}

\newcommand{\ha}{H$\alpha$}
\newcommand{\hb}{H$\beta$}
\newcommand{\mgii}{Mg\,{\sc ii}}

\begin{document}

\title{Reverberation Mapping with Intermediate-Band Photometry:
Detection of Broad-Line \ha\ Time Lags for Quasars at $0.2<z<0.4$}

\author{Linhua Jiang\altaffilmark{1}, Yue Shen\altaffilmark{2,3}, 
Ian D. McGreer\altaffilmark{4}, Xiaohui Fan\altaffilmark{4}, 
Eric Morganson\altaffilmark{3,5}, and Rogier A. Windhorst\altaffilmark{6}
}

\altaffiltext{1}{Kavli Institute for Astronomy and Astrophysics, Peking
   University, Beijing 100871, China; jiangKIAA@pku.edu.cn}
\altaffiltext{2}{Department of Astronomy, University of Illinois at 
	Urbana-Champaign, Urbana, IL 61801, USA}
\altaffiltext{3}{National Center for Supercomputing Applications, University
   of Illinois at Urbana-Champaign, Urbana, IL 61801, USA}
\altaffiltext{4}{Steward Observatory, University of Arizona,
   933 North Cherry Avenue, Tucson, AZ 85721, USA}
\altaffiltext{5}{Max-Planck-Institut f\"ur Astronomie, K\"onigstuhl 17, 
	69117 Heidelberg, Germany}
\altaffiltext{6}{School of Earth and Space Exploration, Arizona State
   University, Tempe, AZ 85287, USA}

\begin{abstract}
We present a reverberation mapping (RM) experiment that combines broad- and 
intermediate-band photometry; it is the first such attempt targeting a sample 
of 13 quasars at $0.2<z<0.9$. The quasars were selected to have strong \ha\ or 
\hb\ emission lines that are located in one of three intermediate bands (with 
FWHM around 200 \AA) centered at 8045, 8505, and 9171 \AA. The imaging 
observations were carried out in the intermediate bands and the broad $i$ and
$z$ bands using the prime-focus imager 90Prime on the 2.3m Bok telescope. 
Because of the large ($\sim$1 deg$^2$) field-of-view (FoV) of 90Prime, we 
were able to include the 13 quasars within only five telescope pointings 
or fields. The five fields were repeatedly observed over 20--30 epochs that 
were unevenly distributed over a duration of 5--6 months. The combination of 
the broad- and intermediate-band photometry allows us to derive accurate light 
curves for both optical continuum (from the accretion disk) and line (from the 
broad-line region, or BLR) emission. We detect \ha\ time lags between the 
continuum and line emission in 6 quasars. These quasars are at a relatively 
low redshift range $0.2<z<0.4$. The measured lags are consistent with the 
current BLR size-luminosity relation for \hb\ at $z<0.3$. While this 
experiment appears successful in detecting lags of the bright \ha\ line, 
further investigation is required to see if it can also be applied to the 
fainter \hb\ line for quasars at higher redshifts.
Finally we demonstrate that by using a small telescope with a large FoV, 
intermediate-band photometric RM can be efficiently executed for a large 
sample of quasars at $z>0.2$.

\end{abstract}

\keywords
{Galaxies: active --- quasars: general --- quasars: emission lines --- 
quasars: supermassive black holes}

\section{INTRODUCTION}

AGN reverberation mapping \citep[RM;][]{bla82,peter93} measures the light
travel time (i.e., lags) between different regions of an AGN, most commonly 
the time lag between the UV/optical continuum (from the accretion disk) and 
the broad line (from the broad-line region, or BLR) emitting regions. RM is a 
powerful tool for probing the structure and kinematics of AGN BLRs. RM is used 
to estimate the masses of AGN central supermassive black holes (SMBHs), by 
combining the relation between the BLR size and AGN luminosity (the $R$--$L$ 
relation) with the assumption of virialized motions of clouds in the BLR.
Through application of this method, RM has been established as the primary 
{\em direct} SMBH mass estimation technique for AGN/quasars at $z\le0.1$.

RM campaigns are expensive and time-consuming. They require repeated 
observations of individual targets with sufficient cadence over durations of a 
few months to a few years, depending on the source redshift and luminosity. 
The success rate also relies on other factors, such as whether the variability 
of the target is significant or not during the RM campaign, which is usually 
unpredictable. To date RM experiments have been successful for about 60 
AGN/quasars
\citep[e.g.][]{kas00,kas05,peter02,peter04,peter14a,ben09,ben13,den09,den10,raf11,raf13,bar11,bar15,du14,du15,wang14,hu15}.
A more detailed history of AGN RM experiments is summarized in a few recent 
works \citep[e.g.][]{ben13,ben15,peter14b,shen15a}.

The majority of the above RM work was done with low-redshift AGN at $z\le0.2$. 
Much higher-redshift ($z\ge1$) AGN/quasars have also been tried
\citep[e.g.][]{met06,kas07,tre07}, 
yet the number of the successful detections of time lags is still very small.
Recently the SDSS Reverberation Mapping program \citep[SDSS-RM;][]{shen15a}
has enabled a new method of carrying out RM experiments. The SDSS-RM program 
is a dedicated multi-object RM campaign that simultaneously targeted 849 
quasars in a single 7 deg$^2$ field. Based on the SDSS-RM data, 
\citet{shen15b} have reported their first detections of time lags in
a sample of quasars at $z\ge0.3$.

Traditional RM programs use spectroscopic observations to monitor the
variability of continuum and line emission. Recently, photometric RM has 
been proposed or performed \citep[e.g.][]{haas11,zu13,che12,che14}.
The basic idea is to take photometry in two bandpasses, with one bandpass
`on' an emission line and the other one `off' the line. The combination of the 
two measurements is used to derive the continuum and line fluxes. The 
advantage of the photometric RM is that it does not require spectroscopic 
observations, and can be easily performed with small telescopes. The 
challenge is the small contribution of the emission line flux to the total
bandpass flux within a broad band, so that the photometric uncertainties 
significantly hamper measurements of variability in the line fluxes. 
Alternatively, one may use a narrow band with a full width at half maximum 
(FWHM) of a few tens \AA\ (up to $\sim$120 \AA) to cover the emission 
line. This has been done for a few local AGN at $z<0.05$  
\citep[e.g.][]{haas11,ram13,poz15}. In this case, the line flux contributes 
a large fraction of the total flux in the narrow band and line variability 
is more readily detected.

In this paper we present our intermediate-band reverberation mapping (IBRM) 
project, which uses the combination of broad and intermediate bands (with FWHM 
around 200 \AA) to perform photometric RM. The usage of intermediate bands has 
the following two advantages, in addition to the general advantages of 
photometric RM mentioned above. An intermediate band can usually cover a
whole emission line, while the line flux still contributes a significant 
fraction of the total flux in the band, if the line is selected to have high 
equivalent width (EW) as we do for the IBRM program.
Secondly, it has a larger (compared to narrow bands) dynamic range in 
wavelength that allows the inclusion of more than one targets per field, 
which substantially increases observing efficiency.
In our IBRM program, we observed 13 quasars within five fields or telescope
pointings, and successfully detected time lags in 6 of them.

The structure of the paper is as follows.
In Section 2 we present our quasar sample selection and their optical spectra. 
In Section 3 we introduce our IBRM campaign and the details of observations
and data reduction. We derive the light curves and time lags of the quasars
in Section 4, and put them in the context of the $R$--$L$ relation in Section 
5. In Section 6 we summarize the paper.
Throughout the paper we use a $\Lambda$-dominated flat cosmology with $H_0=70$
km s$^{-1}$ Mpc$^{-1}$, $\Omega_{m}=0.3$, and $\Omega_{\Lambda}=0.7$.
All magnitudes are on the AB system.

\section{QUASAR SAMPLE}

In this section we present the selection of our quasar sample.
Before we go into the detailed steps, we briefly introduce the telescope and
instrument that we used for the IBRM project, which is directly related to our 
sample selection. The telescope that we used is the Steward Observatory 2.3m 
Bok telescope, and the instrument is its prime focus imager 90Prime.  
90Prime has a large, square field-of-view (FoV) of roughly one degree on a 
side. It uses four 4K thin CCDs that were optimized for U-band imaging 
\citep{zou15}. We used two broad-band filters $i$ and $z$, and three 
intermediate-band filters, BACT12, BACT13, and BACT14. These intermediate-band 
filters were originally designed for the Beijing-Arizona-Taipei-Connecticut 
(BATC) Color Survey \citep[e.g.][]{fan96,yan00,zhou01}. 
The broad-band filters are used to measure 
continuum flux. The effective wavelengths of the three intermediate-band 
filters are 8045, 8505, and 9171 \AA, with the FWHMs of 230, 180, and 264 \AA, 
respectively. They cover three wavelength ranges with relatively weak OH sky 
emission, so imaging in these bands is very efficient.

\subsection{Sample Selection}

Our sample selection began with the SDSS DR7 quasar catalog delivered by
\citet{sch10} and \citet{shen11}. The emission lines used for our IBRM project
are \ha\ and \hb, two of the strongest lines in quasar spectra. 
We first selected quasars at certain redshifts so that their \ha\ or \hb\ 
emission lines are located in one of the three intermediate bands (the center 
of an emission line is roughly within the central 50\% of the filter.)
Specifically, the redshift ranges considered here are [0.216, 0.236], 
[0.290, 0.302], [0.385, 0.410], [0.642, 0.668], [0.741, 0.758], and 
[0.870, 0.904], and there are 4326 quasars in these redshift ranges.  
We then selected quasars in a certain coordinate range, because the 
observations of our IBRM project shared the Bok nights with the SDSS-RM 
project \citep{shen15a}, as we will see in the next section. The coordinate
range chosen here is 8h$<$R.A.$<$13h and Decl.$>$25 Deg, and 1227 quasars 
passed this selection. We further selected targets in a certain brightness 
range (namely, $17<i<19.5$ mag) with high \ha\ or \hb\ EW. 
The \ha\ and \hb\ EW values were measured from the SDSS spectra and 
taken from \citet{shen11}. We required that
the observed \ha\ EW was greater than 180 \AA, or the observed \hb\ EW was
greater than 90 \AA. This ensures that the line emission contributes a
significant fraction of the total flux in the intermediate bands. This is
one of the keys for the success of this program. The choice of $i<19.5$ mag 
was to ensure that we can get high SNRs in the intermediate bands with 5 min 
integration time. The choice of $i>17$ mag was to select quasars with expected 
time lags (in the observed frame) shorter than the duration of our observing 
campaign (roughly 5--6 months). The expected time lags for most of the 
selected quasars are between 20 and 60 days. The observed-frame time lags 
also depends on redshift due to the time dilution of $1+z$ and the strong 
dependence of the intrinsic luminosity on redshift for a given apparent 
magnitude. Therefore, for very bright quasars at relatively high redshifts
($z\ge0.6$), their expected time lags could be significantly longer and
even close to the duration of our RM campaign.
We selected 622 quasars in this step.

\begin{deluxetable*}{cccccccc}
\tablecaption{Quasar Sample}
\tablehead{\colhead{ID} & \colhead{R.A. (J2000)} & \colhead{Decl. (J2000)} &
   \colhead{Redshift} & \colhead{$i$ (mag)} & \colhead{Line} & \colhead{EW} &
	\colhead{Filter}}
\startdata
F1a & 09:00:45.293 & +33:54:22.38 &   0.228 &  17.93 & \ha & 205 & BATC12 \\
F1b & 09:01:56.250 & +33:33:49.49 &   0.878 &  19.00 & \hb & 71  & BATC14 \\
\tableline
F2a & 09:46:59.593 & +29:32:51.13 &   0.387 &  18.58 & \ha & 606 & BATC14 \\
F2b & 09:50:46.582 & +29:38:26.90 &   0.233 &  18.20 & \ha & 299 & BATC12 \\
\tableline
F3a & 11:27:59.260 & +36:02:07.00 &   0.667 &  18.42 & \hb & 132 & BATC12 \\
F3b & 11:29:56.532 & +36:49:19.24 &   0.398 &  19.05 & \ha & 563 & BATC14 \\
F3c & 11:31:14.956 & +36:02:38.30 &   0.229 &  18.22 & \ha & 263 & BATC12 \\
\tableline
F4a & 11:45:53.152 & +28:13:13.41 &   0.401 &  18.73 & \ha & 230 & BATC14 \\
F4b & 11:46:34.914 & +28:26:41.96 &   0.225 &  18.11 & \ha & 249 & BATC12 \\
F4c & 11:49:36.368 & +27:44:04.83 &   0.748 &  18.95 & \hb & 192 & BATC13 \\
\tableline
F5a & 12:36:35.259 & +45:02:08.06 &   0.401 &  18.74 & \ha & 353 & BATC14 \\
F5b & 12:36:58.110 & +45:53:54.46 &   0.235 &  18.12 & \ha & 433 & BATC12 \\
F5c & 12:38:42.730 & +45:18:24.73 &   0.229 &  17.34 & \ha & 292 & BATC12 \\
\enddata
\tablecomments{The table lines separate five different fields. 
Column 7 shows the observed-frame EW values in units of \AA.}
\end{deluxetable*}

After we obtained the list of the quasars from the above steps, we chose the
area/fields that have more than one quasar per square degree (the FoV of the
90Prime). This was to increase the efficiency of the project. Meanwhile, we
matched the quasars to the Pan-STARRS1 \citep[PS1;][]{sch12,ton12,mag13}
preliminary catalog, and obtained their variability values as follows.
For each quasar, we extracted the standard deviations of the magnitudes from 
the catalog. There are five standard deviation values for five PS1 bands
($g_{\rm P1}$, $r_{\rm P1}$, $i_{\rm P1}$, $z_{\rm P1}$, and $y_{\rm P1}$), 
and we took the average of the second and third largest standard deviation 
values as the variability of this quasar. We then eliminated $\sim$10\% of the
sources whose variability was roughly consistent with error bars. Finally,
from the remaining sources we selected 13 quasars or
5 fields for the IBRM project, by considering the following: 1) the fields
are roughly evenly distributed between 8h and 13h for the convenience of
observations; 2) the quasars have relatively strong \ha\ (or \hb) EWs
(stronger are better); 3) the quasars show relatively strong variability.
When the demands for 2) and 3) are difficult to meet simultaneously,
we slightly favored 2), because quasar variability is a stochastic process.

Table 1 lists the details of the 13 quasars.
Column 1 shows the ID of the quasars. We use `F1' to `F5' to denote the five
fields, and use `a' to `c' to denote the quasars within a field. The following
four columns are the coordinates, redshifts, and $i$-band magnitudes drawn
from the SDSS DR7 catalog. Columns 6 and 7 shows the emission lines that we
used and their EWs in the observed frame. Column 8 shows the intermediate
filters that cover the lines. These quasars covers a redshift range of
$0.22<z<0.88$. We mainly use the \ha\ line (10 out of 13
quasars), because the \ha\ line is much stronger than the \hb\ line, and
thus quasars with strong \ha\ were preferentially selected.
Note that the EW of F1b is lower than
our selection criterion. This is because the line is close to the red end of
its SDSS spectrum, and the EW value given by the SDSS DR7 is not accurate.
The value given in the table was measured from an MMT spectrum with much
better quality (see the next subsection). The contributions of line emission 
to the intermediate-band fluxes are roughly between 35\% and 70\% for all
objects except F1b, for which the line contribution is only $\sim$20\%.

\subsection{Optical Spectra}

Single-epoch quasar optical spectra are needed to accurately derive the line 
contribution to the broad-band photometry. 
These quasars have optical spectra from SDSS I and 
II. The SDSS spectra do not cover the wavelength range beyond $\sim$9100 \AA,
which is needed for the BATC14 filter. For the quasars observed with the
BATC14 filter (see Table 1), we obtained new optical spectra using the MMT 
Red Channel spectrograph. The observations were carried out as backup targets 
during other programs in 2014, when weather conditions were poor. The 
observations were made in long-slit mode with a spectral resolution of 
$\sim$10 \AA. 
The integration time was roughly 5--10 min per object, which is sufficient for 
our purposes. The spectra were reduced using standard IRAF 
routines\footnote{IRAF
is distributed by the National Optical Astronomy Observatory, which is
operated by the Association of Universities for Research in Astronomy (AURA)
under cooperative agreement with the National Science Foundation.}.
The MMT spectra cover a wavelength range of 7000--10000\AA. 
The final spectra of these quasars are the combination of the SDSS and MMT
spectra.

Figure 1 shows the optical spectra of our quasar sample in the observed frame.
As we mentioned above, some spectra were directly taken from the SDSS, while
the others were the combination of the SDSS and MMT spectra. In each panel of 
Figure 1, we also show the transmission curves of the 90Prime $i$ 
and $z$ filters (the blue dotted profiles) and one of the intermediate filters 
(the red dotted profile) that covers \ha\ or \hb. The CCD quantum efficiency
has been taken into account. Note that the transmission curves of the 90Prime 
$i$ and $z$ filters are slightly different from those of the SDSS $i$ and $z$ 
filters.

In several cases, the intermediate bands do not entirely cover the 
emission lines (e.g. F2a and F5b). The effect of missing line wings on the 
measurement of BLR sizes is very small. When we calculate line emission for 
light curves in section 4, we only consider the contribution from the part 
covered by the intermediate bands, which contains more than 90\% (in most 
cases more than 95\%) of the total line flux. \citet{poz14} conducted detailed 
calculations to estimate the consequences of the above missing line wings on 
photometric RM results, and concluded that the effect on the measurements of
BLR sizes is only a few per cent. This is negligible compared to the size
measurement uncertainties we will get in section 4.

\begin{figure*} 
\epsscale{1.0}
\plotone{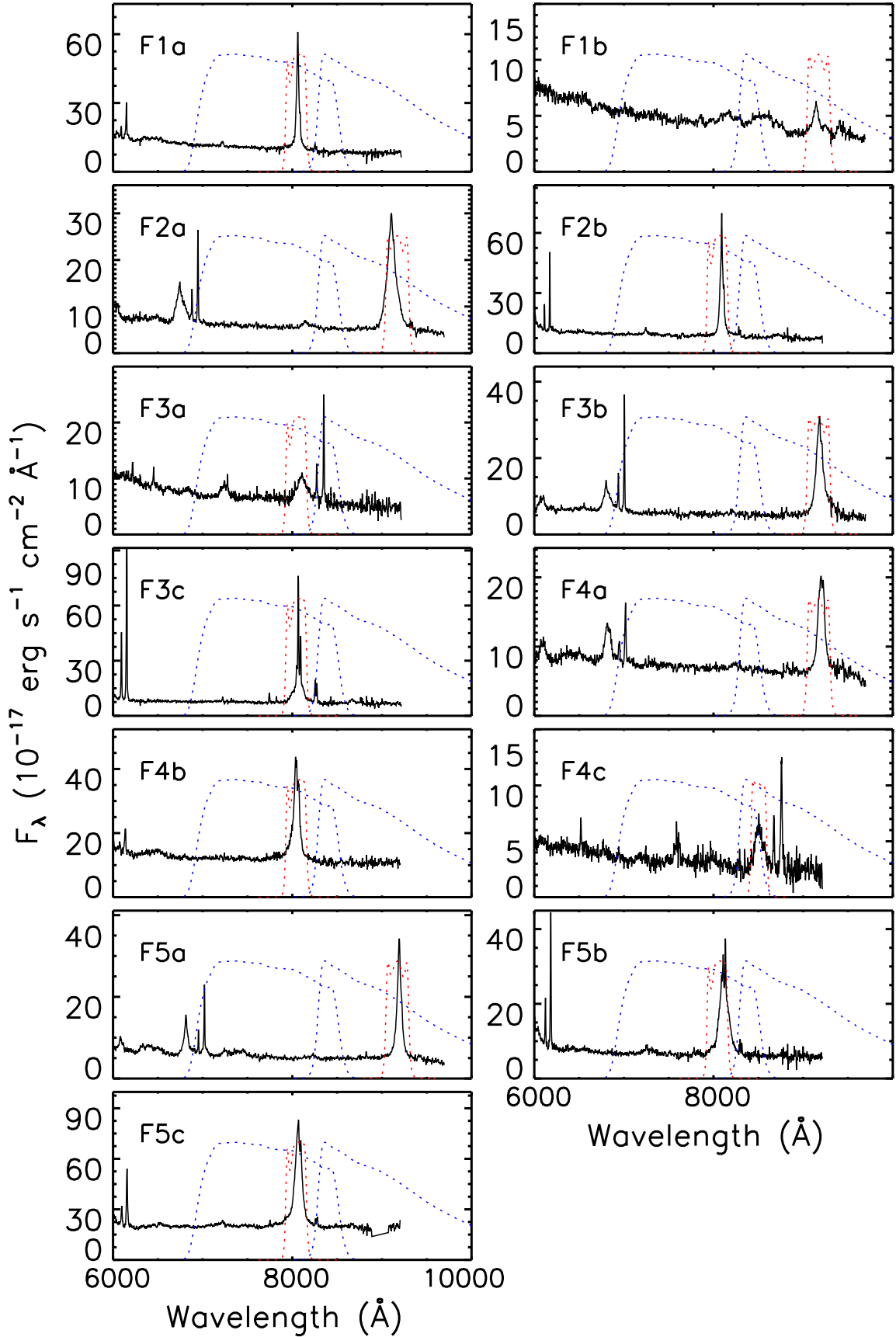}
\caption{Observed-frame optical spectra of the 13 quasars in our sample.
The spectra were taken from the SDSS and the MMT. The blue dotted profiles
represent the transmission curves of the 90Prime $i$ and $z$-band filters, and 
the red dotted profiles represent three intermediate filters that covers \ha\ 
or \hb. The CCD quantum efficiency has been taken into account, and
the curves have been normalized so that their peak values are the same.}
\end{figure*}

\section{OBSERVATIONS AND DATA REDUCTION}

The IBRM project was carried out in the spring semester, 2014. 
It shared the Bok nights with the SDSS-RM project. As we mentioned earlier, 
the SDSS-RM project was one of the SDSS III ancillary projects.
It used the CFHT and Bok telescopes to do broad-band ($g$ and $i$) photometry
for measurements of continuum light curves in SDSS-RM \citep{shen15a}.
Our targets and observing time were coordinated with the SDSS-RM
project.

\subsection{Bok Observations}

The Bok observations of the quasars were conducted in January through June, 
2014. Due to the constraints from the telescope scheduling, we obtained one or 
two long observing blocks each month (see section 4.1). Hence the Bok nights 
for the SDSS-RM and IBRM projects were not evenly distributed, and were 
clustered around the nights with relatively bright moon phase. Such an 
observing schedule does not provide an optimal cadence for RM studies.
Each of the five fields were observed for between 20 and 30 epochs over the
full campaign.

Because of the large FoV of the 90Prime, the 13 quasars in our sample were 
covered by only 5 telescope pointings or fields from F1 to F5. 
We usually observed at least 3 fields per epoch/night. 
The observations were made via observing scripts. Each time after we slewed 
the telescope to a new field, the scripts automatically changed filters, 
tweaked focus, and took the data, in the order of $i$, $z$, BATC12, BATC14 
(and BATC13 for Field 4). The typical on-source integration time was 150 s 
in the $i$ band, and 300 s in the other bands.
The observing conditions were mostly moderate with clear skies, moderate
seeing ($\sim1.5\arcsec$), but significant moonlight.

\subsection{Data Reduction}

The 90Prime images were reduced in a standard fashion using our own {\tt IDL}
routines. The basic procedure was described in \citet{jiang15}.
First, we made a master bias image and a master flat image from bias and flat 
images taken in the same night. A bad pixel mask was also created from the 
flat image. Science images were then overscan and bias-corrected and 
flat-fielded. Next we identified saturated pixels and bleeding trails, and
incorporated them (along with the bad-pixel mask) into the weight images. 
The affected pixels were interpolated over in the science images.
We call the science images at this stage `corrected images'.

The 90Prime CCDs are thin chips, and thus produce strong fringing in the bands
that we used. We subtracted sky background and fringes using two iterations. 
The first-round of sky subtraction was
performed by fitting a low-order 2D polynomial function to the background.
A master fringing image (per filter) was made by median stacking at least
eight sky-subtracted images in the same filter. This fringing image was scaled
and subtracted from the original `corrected' images (before sky-subtraction 
was done). We detected objects in the fringe-subtracted images using 
{\tt SExtractor} \citep{ber96}. A better sky image was produced from each 
fringe-subtracted image with the detected objects masked out. 
Then the second round of sky and fringe subtraction was performed, but with 
the detected objects masked out.

In order to derive astrometry, we detected bright objects using 
{\tt SExtractor} \citep{ber96}, and calculated
astrometric solutions using {\tt SCAMP} \citep{ber06} by matching objects to
the SDSS. With the new astrometry we re-mapped the images using {\tt SWARP} 
\citep{ber02}. The re-mapped images have a native pixel size of 
$0.455\arcsec$.

\subsection{Photometry}

Accurate (relative) photometry is another key for the success of this project.
In order to achieve accurate photometry, for any given quasar in the whole
observing campaign, we used a large number of nearby bright point 
sources for photometric calibration. These bright sources and our 
quasar targets was always located in roughly the same part of the same CCD, 
which minimizes the effect from any large-scale systematics.
This allows us to achieve relatively small 
uncertainties ($\sim0.01$ mag) on the magnitude zero points that are 
usually negligible compared to the uncertainties in the light curves that
we derive in Section 4. The details are as follows.

The four CCDs were read out via 16 amplifiers, with 4 amplifiers per CCD.
For any of the 13 quasars in an image, we only performed photometry for the
amplifier area in which this quasar was located (roughly $15\arcmin$ on a 
side). We did not use other parts of the image (for this quasar) due to the 
possible small zero point shift across the amplifiers and CCDs \citep{zou15}. 

We first chose a `standard' night, which was photometric and relatively dark,
and performed photometry for the images taken in this standard night.
Photometry was measured within an aperture (diameter) size of 8 pixels 
($\sim3\farcs6$) using {\tt SExtractor}. 
We then picked up bright (at least 30$\sigma$ detection)
but unsaturated point sources, and matched them to the SDSS. The density of 
the bright stars is about 50--100 per amplifier. We used the SDSS PSF 
magnitudes, so the measured magnitudes are total magnitudes with
aperture corrections automatically taken into account.
After we obtained the $i$ and $z$-band magnitudes for a given object, we
calculated its intermediate-band magnitudes as follows. We assumed that its
spectrum in the wavelength range of the $i$ and $z$ bands (also covers the 
three intermediate bands) was a power law, which is determined by its $i$ and 
$z$-band magnitudes. Then its intermediate-band magnitudes were directly 
calculated from this power-law spectrum and the intermediate-band transmission 
curves. The resultant magnitudes have very weak dependence (usually 
$\Delta m \le0.03$ mag) on the assumption of the spectrum shape, as long as
there are no emission or absorption lines in the intermediate bands.
Such small uncertainties on the zero magnitude points have no effect on our
measurement of light curves, which reply on relative photometry.
We calculated AB magnitudes for all these 
bright stars taken in the standard night. These bright stars were used 
as standard stars for all other images.

We then measured photometry for the images taken in all other nights.
The procedure was the same. But we used the bright stars found in the standard
night as `standard stars', and used their magnitudes for absolute flux 
calibration. Because of the large numbers of bright standard stars for any
given quasar, we achieved high accuracy on relative zero flux points.
Figure 2 illustrates this point. Its horizontal axis shows the magnitude 
difference of the standard stars taken between the standard night and the
first several nights other than the standard night. 
The vertical axis shows the distributions of the magnitude difference.
The sigma values quoted in the figure were estimated by fitting a Gaussian 
profile to the top 80\% of the distributions. 
The bottom 20\% deviates from the Gaussian distribution, likely due to 
some unreliable `standard' stars, such as variable stars. 
The tight distributions of the magnitude difference in the four bands suggest 
that our accurate flux calibration is about 0.010 mag rms. We did not plot the 
distribution in BATC13, which is similar to those in the other 4 bands.

\begin{figure} 
\epsscale{1.1}
\plotone{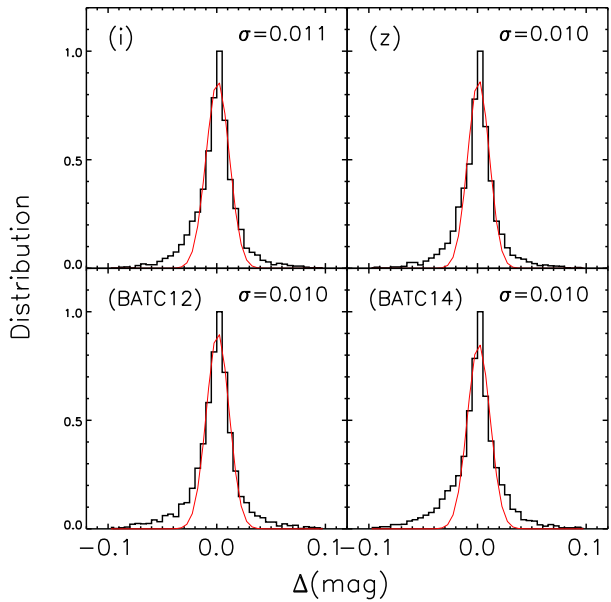}
\caption{Accuracy of the flux calibration.
The horizontal axis shows the magnitude difference of the standard stars taken
between the standard night and the first several nights other than the
standard night. The vertical axis shows the normalized distribution of the
magnitude difference. The standard deviation ($\sigma$) values were estimated
by fitting a Gaussian profile (the red profiles) to the top 80\% of the
distributions (the bottom 20\% significantly deviates from the Gaussian
distribution). The tight distributions in the four bands suggest that our flux
calibration is very accurate.}
\end{figure}

Figure 3 shows the measurement uncertainties as a function of total magnitude
in 4 bands for all quasars observed in the whole RM campaign. The errors 
are estimated within an aperture (diameter) size of 8 pixels from 
{\tt SExtractor}. The uncertainties 
from the absolute flux calibration are not included in this plot.
In our images, the noise is completely dominated by sky background, so these 
errors are quite reliable (background variance reflects errors). 
The errors are mostly smaller than 0.02 mag. In rare cases errors can be 
larger than 0.05 mag, mostly caused by low sky transparency.

The final photometric errors take into account (quadratically) the 
measurement errors (Figure 3), the errors from the flux calibration 
(Figure 2), and the uncertainties due to varying seeing. A quasar host galaxy
is not a point source, and its radial profile is broader than PSF, so the
aperture correction derived from point sources does not precisely correct for 
all light loss. This results in photometric variation with varying seeing.
We estimate this variation as follows. In section 5.1, when we measure the
host galaxy contribution for each quasar, we generate a combined image 
(from single-epoch images with good seeing), build a PSF model image, and 
derive a host-galaxy image using image decomposition. We make use of these
`deep' and model images, because single-epoch images do not have sufficient 
SNR. We convolve these images with Gaussian kernels, which mimics varying 
seeing. We then perform aperture photometry in the same manner as in 
single-epoch images. The only difference is that the aperture 
correction is measured from the corresponding PSF images. We find that 
when PSF varies from $\sim1.2\arcsec$ to $\sim1.8\arcsec$ (almost covers our 
seeing range), the photometric variation is about $0.007\pm0.002$ mag.
Such variation is small for our targets, partly because we used a large
aperture for photometry. In many cases, however, it is comparable to the
measurement errors shown in Figure 3. Thus we carry this error of
0.007 mag to the final photometric errors.

\begin{figure} 
\epsscale{1.1}
\plotone{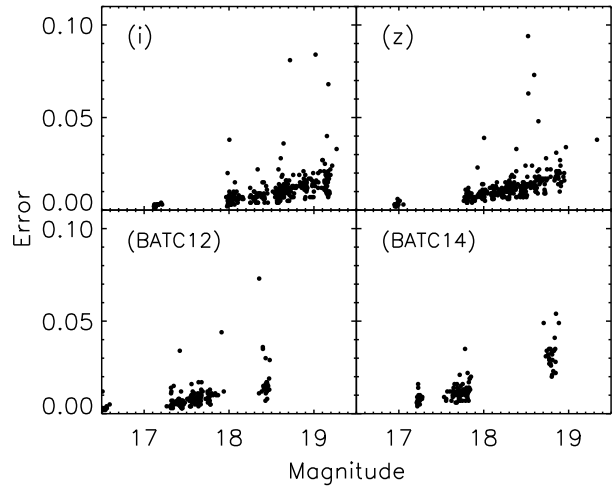}
\caption{Measurement uncertainties as a function of total magnitude
for all quasars observed in the whole RM campaign. The errors
are measured within an aperture (diameter) size of 8 pixels. The
uncertainties from absolute flux calibration are not included.
The errors are mostly smaller than 0.02 mag. In rare cases,
errors can be larger than 0.05 mag caused by low sky transparency.}
\end{figure}

\section{TIME LAGS}

In this section we present our main results. 
For each quasar in Table 1, we first compute the continuum and emission line
light curves. This step is straightforward, as we have decent 
optical spectra and accurate broad- and intermediate-band photometry. We 
then derive the time lag between continuum and line emission in standard ways.

\subsection{Light Curves}

\begin{figure*} 
\epsscale{1.0}
\plotone{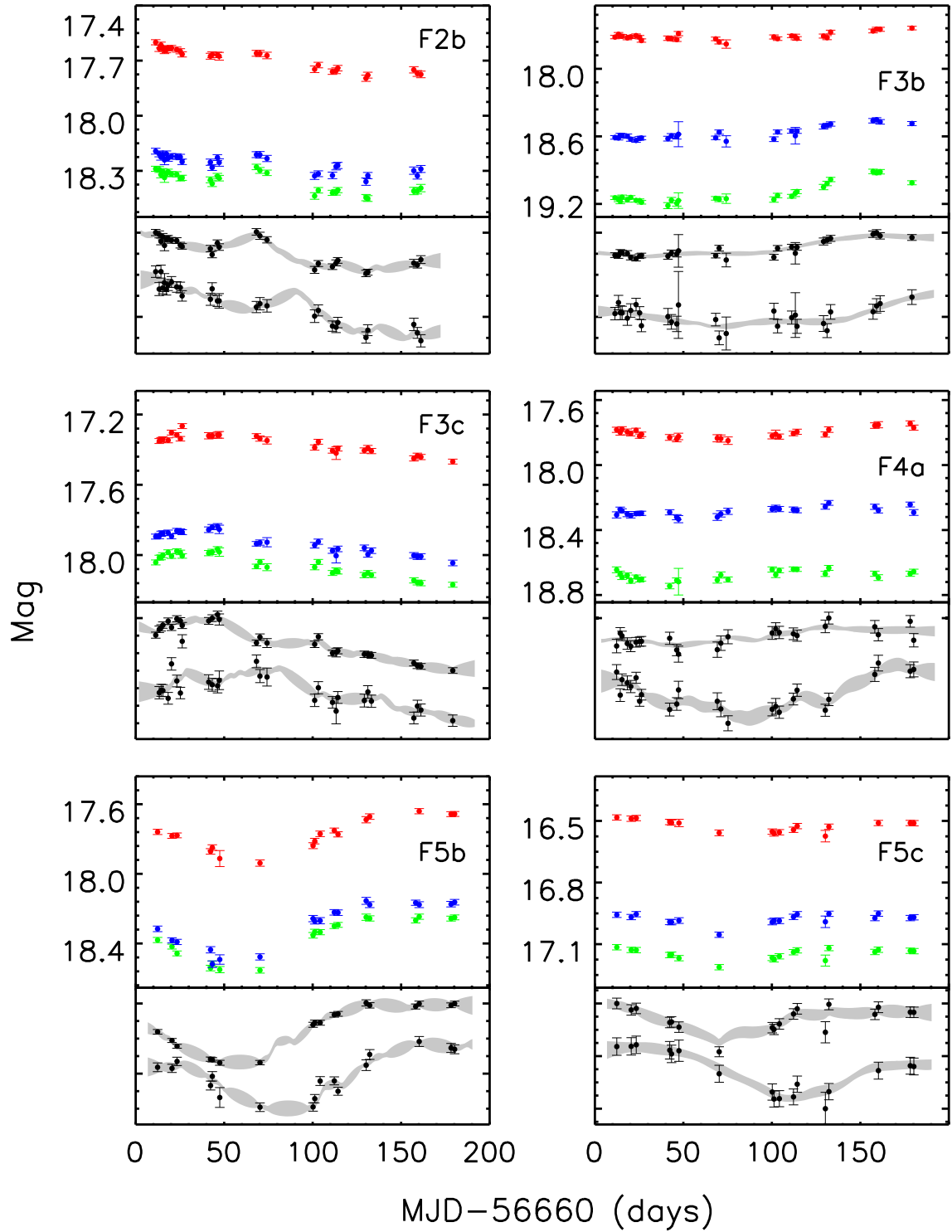}
\caption{Light curves of the 6 quasars that show significant time lags during 
our IBRM campaign. For each object in the upper panel, the blue and green 
circles show the light curves in the $i$ and $z$ bands, and the red circles 
indicate the light curve in the intermediate band. The lower panel shows the 
light curves of the continuum flux (upper curves) and line flux (lower 
curves). See section 4.1 for details on how the line and continuum flux is 
derived. These two curves have been shifted along the y axis so that
they are displayed clearly. The gray shaded curves indicate the simulated 
light curves ($1\sigma$ area) as computed from JAVELIN (section 4.1).}
\end{figure*}

For each quasar we have a single-epoch optical spectrum and a series of $i$, 
$z$, and intermediate-band photometric measurements. We first calculate 
the contribution of the line emission (or equivalently the contribution of the 
continuum emission) to the broad-band photometry using the optical spectrum. 
We select regions with
little line emission in the spectrum as continuum windows, and fit a 
power-law curve ($f_\lambda=b\times \lambda^{\alpha}$) to these windows. 
This power-law continuum may contain a central AGN component and a host
galaxy component (see the next section). As long as the host galaxy does not
vary (a constant component), the inclusion of the host galaxy component 
does not affect the determination of time lags.
The line emission is obtained by subtracting the power-law continuum
from the spectrum. We assume that the contribution of the line emission to
the broad-band photometry does not vary with time. The reason is that the
line contribution is smaller than 5\%, and the line variability is usually
smaller than 20\%, so the effect of line variability on broad-band photometry
is smaller than 1\%. As we will see below, the continuum value we derive
for a quasar is determined by two broad bands ($i$ and $z$), 
with one band without line contamination, so the effect of line variability on 
continuum is even smaller by roughly a factor of two, which is much smaller 
than the uncertainties in the light curves derived below. Thus our above 
assumption has a negligible effect on the measurement of the continuum, 
but largely simplifies our procedure.

We then derive the line flux and continuum flux at the wavelength of
the intermediate band from their corresponding $i$, $z$, and intermediate-band
photometry at each epoch. The continuum components at $i$ and $z$ are
computed by subtracting the line contribution from these bands. 
A power-law continuum 
is derived analytically from the two flux measurements at the effective
wavelengths of the $i$ and $z$ bands. We then determine the continuum value
at the effective wavelength of the intermediate-band from the power-law
continuum. This continuum value is the continuum that will be used for light 
curves. Finally the line emission is obtained
by subtracting the continuum component from the intermediate-band photometry.
Figure 4 plots the light curves of the 6 quasars that show significant time 
lags during our IBRM campaign (see the next subsection). 
For each object, the upper panel shows the light curves in the $i$
(blue circles), $z$ (green circles), and one of the intermediate bands 
(red circles). The lower panel shows the light curves of the continuum flux 
(upper curves) and line flux (lower curves).

\subsection{Lag Measurements}

\begin{figure} 
\epsscale{1.1}
\plotone{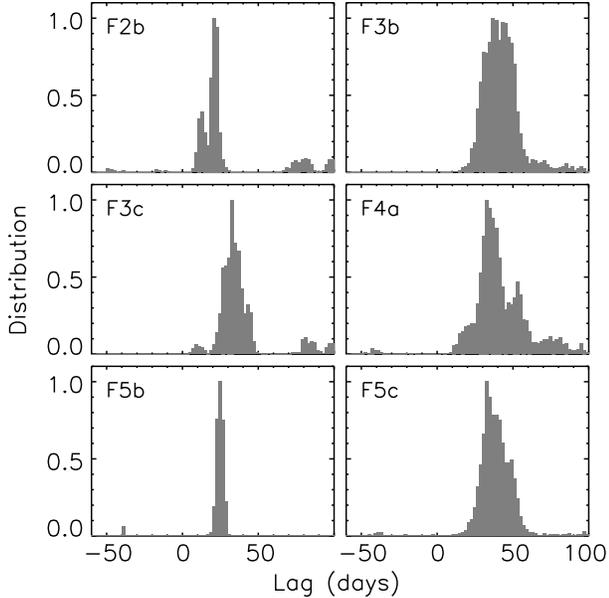}
\caption{Time lags (in the observed frame) for 6 quasars with significant
lag detections.
The lag distribution for each quasar is based on 10,000 experiments using
JAVELIN \citep{zu11}. See section 4.2 for details. The distributions have been
normalized so that the peak values are equal to 1.}
\end{figure}

\begin{deluxetable}{cccccccc}
\tablecaption{Time Lags and Luminosities of the 6 Quasars}
\tablehead{\colhead{ID} & \colhead{Redshift} & \colhead{Lag (days)} &
   \colhead{Log($\lambda L_{\lambda}$)} & \colhead{$f_{\rm host}$} }
\startdata
F2b &  0.233 &  $21.1_{- 8.3}^{+46.9}$  &  43.84 & 0.21 \\
F3b &  0.398 &  $41.2_{- 9.9}^{+ 9.8}$  &  44.17 & 0.19 \\
F3c &  0.229 &  $34.0_{- 6.5}^{+ 9.7}$  &  43.84 & 0.37 \\
F4a &  0.401 &  $38.1_{- 9.1}^{+17.8}$  &  44.32 & 0.28 \\
F5b &  0.235 &  $25.0_{- 2.3}^{+ 2.2}$  &  43.79 & 0.22 \\
F5c &  0.229 &  $37.9_{- 7.4}^{+10.3}$  &  44.25 & 0.36 \\
\enddata
\tablecomments{The time lags are in the observed frame. The quasar 
luminosities (in units of erg\,s$^{-1}$) have been corrected for host-galaxy 
contamination. The last column shows the contribution from host galaxies
(section 5.1).}
\end{deluxetable}

We estimate time lags between the line and continuum emission derived 
above using the JAVELIN package \citep{zu11,zu13}. As we have seen, the light 
curves were unevenly (sometimes sparsely) sampled, due to the constraints from 
telescope time scheduling. JAVELIN is able to deal with such an uneven
sampling. It assumes that the variability of a quasar/AGN can be well 
described by the damped random walk (DRW) model \citep[e.g.][]{kelly09}, 
and its emission line light curve is the lagged and scaled version of its 
continuum light curve. For a given quasar, we first model its continuum 
variability using JAVELIN, and find the distribution of the DRW parameters. 
We then statistically interpolate the continuum light curve. The light curve 
is shifted, smoothed, and scaled, before it is compared to the corresponding 
emission-line curve. 
The smoothing here refers to the use of a transfer function 
(non-Delta-function ) in JAVELIN to mimic the realistic line response to 
continuum light curves. This is due to the fact that the BLR clouds are 
distributed at different radii with different velocities, which results in a 
transfer function that is broad in lag. 
For details, see the JAVELIN papers \citep{zu11,zu13}.
This process is performed 10,000 times using the MCMC method. 
The final results are the best model fits for each try. 
Note that all calculations above are based on flux (not magnitudes).

Based on the time lag measurements from JAVELIN, we find that 6 (out of 
13) quasars in our sample show clear lag detections during our IBRM campaign 
(we will discuss the other quasars in the next subsection). 
The lag range allowed in the above 
calculation is from --100 to +100 days. We do not consider a wider range, 
simply because the duration of the IBRM campaign was only 150--170 days.
Figure 5 shows the distributions of the measured lags for the 6 quasars (their 
light curves are shown in Figure 4). They show clear single distribution peaks.
The results are listed in Table 2. The lag errors in the table are calculated 
by including 16\% and 84\% of the total distributions 
around the median distribution (50\% of the total). 
Obviously they depend on the lag range that we consider.
On the other hand, they are not sensitive to the lag range, as long as the lag
detections are substantial with most distributions clustered around the 
median values. This applies to all quasars except F2b in Figure 5.
The lag measured for F2b is $21_{-8}^{+47}$ days. It has a large upper error
due to the non-negligible fraction of the total distribution beyond 70 days.
If we were to reduce the lag range to $-100\sim70$, the lag becomes
$21_{-7}^{+2}$ days, with a much smaller upper error. 
We adopt the larger error for consistency in the paper.

We further use the discrete correlation function (DCF) to validate the above
lag detections. The algorithm we adopt is the $z$-transformed DCF (zDCF), 
which was designed to handle unevenly sampled light curves \citep{ale13}.
The estimated DCFs for the 6 quasars are shown in Figure 6. These quasars 
also show clear DCF peaks. The lag uncertainties from the zDCF are larger. One 
reason is that the zDCF uses time-lag bins, and the minimum number required in 
each bin is roughly 11 for meaningful statistics in the zDCF. Given the small
numbers of epochs in our IBRM project, the zDCF can only coarsely sample 
the light curves in the lag time space. Nevertheless, the zDCF peaks in Figure 
6 are consistent with the results from JAVELIN except for F2b.
F2b shows a significant lag detection by JAVELIN in Figure 5, but its zDCF 
peaks at $\sim0$. The lag detection from JAVELIN could be real, or a false 
positive because of the sparse-sampling in light curves. On the other hand,
the zDCF of F2b is broad, which is not against a lag of $\sim20$ days.
This can be solved with more evenly and finely sampled 
light curves in the future.

We perform a simple experiment to test our results.
For any pair of light curves (continuum and line) in the
6 quasars, we randomly re-order one curve, and repeat the above processes using
JAVELIN to estimate the rate of false positives. This is done a hundred times
for each quasar (each pair of light curves). These tests generally show
small false positive rates. For the first four quasars that were observed in
nearly 30 epochs, the false positive rates are only 3--4\%. The rates increase
to 7--9\% for the last two quasars that were observed only $\sim$20 times
(see \citet{shen15b} for a detailed discussion).

\begin{figure} 
\epsscale{1.1}
\plotone{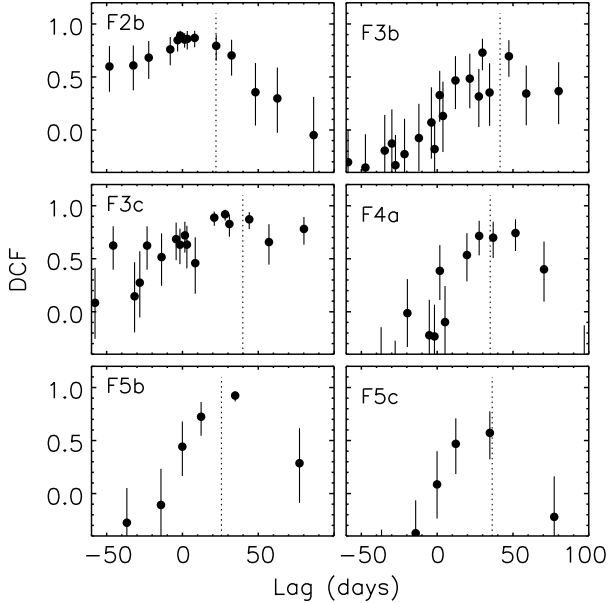}
\caption{DCFs for the 6 quasars with significant lag detections (in the
observed frame). The DCFs are estimated using the zDCF \citep{ale13}.
The dotted vertical lines indicate the lags measured from JAVELIN (Table 2
or Figure 5). The results from the zDCF and JAVELIN are consistent.}
\end{figure}

The detected lags in the 6 quasars are all for \ha, and at relatively low
redshifts between 0.22 and 0.40. It is not surprising, because 10 out of the 13
lines in the original sample are \ha, and \ha\ is much stronger than \hb\
on average. In addition, higher redshifts usually mean higher intrinsic
luminosities and larger time dilution $(1+z)$, leading to much larger observed
time lags that are likely beyond the detection capability of our IBRM campaign.

\subsection{Quasars without Significant Lag Detections}

We did not detect time lags between continuum and line emission in the other
7 quasars in our sample. Similar to Figure 6, Figure 7 shows the distributions
of the lags for these quasars from JAVELIN. Unlike those in Figure 6, 
quasars in Figure 7 do not show single strong peaks. They rather show 
multiple peaks or continuous distributions. 
There are two main reasons for these non-detection.
The first reason is that the expected time lags based on the current $R$--$L$
relation \citep[e.g.][]{ben13} are comparable to the duration
of our campaign. For example, the lags expected for F1b, F3a, and F4c are
greater than 120 days, primarily due to their high redshifts. 
The second reason is that the variability is small, or that 
the light curves are relatively flat. These quasars show moderate to large
variability in the PS1 data. 
But quasar variability is a stochastic process, and past
large variability does not guarantee large variability in the future.
In addition, large gaps in light curves can often cause aliasing, and it 
seems the case for F4c. 
For these objects, our data were insufficient to detect lags.

\begin{figure} 
\epsscale{1.1}
\plotone{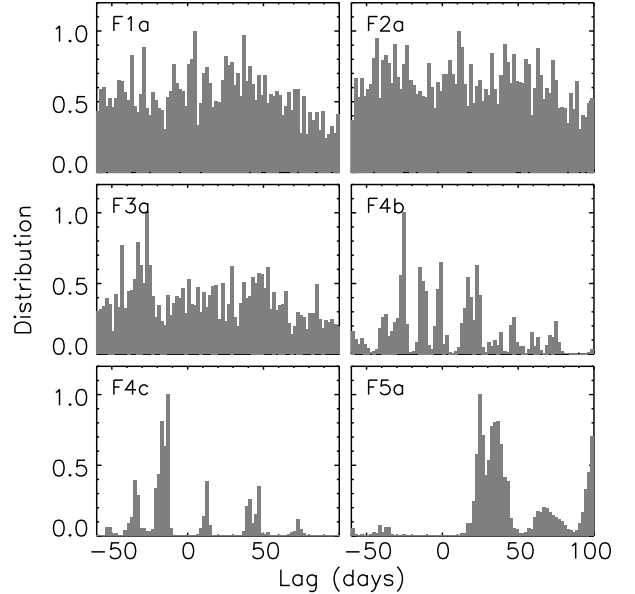}
\caption{Distributions of the time lags derived from JAVELIN for 6 of the 7
quasars without obvious lag detections. The distributions have been
normalized so that the peak values are equal to 1.}
\end{figure}

\section{BLR SIZE-LUMINOSITY RELATION}

\subsection{Light from Host Galaxies}

Quasar host galaxies may contribute a significant fraction of the total light 
in the bands that we measured. There are two general methods to estimate the
light from the hosts: image decomposition and spectral decomposition.
We do not have the high SNR spectra that spectral decomposition requires 
\citep[e.g.][]{shen15c}, so we rely on image decomposition.
Image decomposition works better on images with better PSFs (or seeing). The 
site of the Bok telescope does not deliver good seeing, and the average PSF 
size of our images is about $1.5\arcsec$. In order to construct a deep 
combined image with a decent PSF for each quasar, we choose 50\% of the 
$i$-band images with the best PSF sizes, and co-add them to a stacked image.
The reason to choose the $i$-band is twofold. One is that it is the 
deepest band. The other one is that its effective wavelength is close to
the rest-frame 5100 \AA\ (the commonly used wavelength) for our quasars.
As for flux calibration, we only consider 1/16 of the image, i.e., the 
amplifier that the quasar is located in. This is to avoid any possible large
PSF variation across the large FoV. The PSF sizes of the combined images are 
about $1.1\arcsec$, which is good enough for image decomposition on 
low-redshift quasars \citep[e.g.][]{mat14}.

\begin{figure} 
\epsscale{1.1}
\plotone{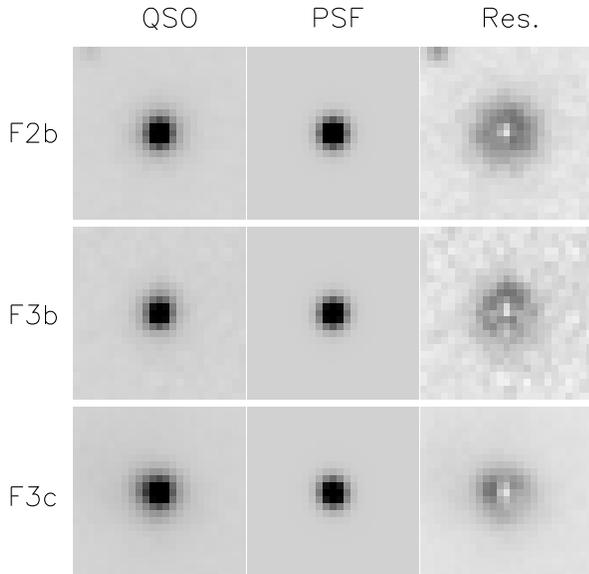}
\caption{Examples for image decomposition. We show the first 3 of
the 6 quasars with lag detection. For each quasar, the first column shows the
combined $i$-band stamp image, with the quasar well centered in the middle.
The middle column shows the PSF image constructed from the combined image.
The third column is the residual image or the host galaxy component after
the PSF image is scaled and subtracted from the quasar image.
The quasar and PSF images have the same intensity scale. The residual images
have a different intensity scale.}
\end{figure}

We perform image decomposition on the combined $i$-band images.
For each quasar/image, the detailed steps are as follows.
We first derive a PSF model for the quasar using {\tt PSFex} \citep{ber11}.
{\tt PSFex} finds point sources in the image, and builds PSF models
as a function of position. We take the PSF model that is closest to the
quasar position as the PSF model for the quasar, although the PSF variations 
are quite small across a single amplifier in our images.
The PSF image size is 25 pixels on a side, with the peak pixel centered in
the middle. We then calculate an accurate central position for the quasar
in the image, and re-sample the image to produce a stamp image centered on 
the quasar. The size of the stamp image is also
$25\times25$ pixels. Following \citet{mat14},
we assume that the central pixel (peak value) of the quasar image is 
completely dominated by the quasar/AGN component. Under this assumption, 
we scale the PSF and subtract it from the quasar image. The residual is 
referred to as the host galaxy component.
Figure 8 demonstrates the procedure by showing three quasars.

From the above procedure, the fraction of the AGN (or host) component is
simply calculated by doing aperture photometry on the PSF image and the quasar
image. In our 6 quasars, the host contribution is roughly between 19\%
and 37\%. In order to measure the luminosity of an AGN at the rest-frame 5100
\AA, we scale its spectrum in Figure 1 to match the mean of the $i$-band
magnitudes obtained in IBRM. We then calculate the AGN luminosity from the
spectrum after removing the host contribution.
The absolute values of the AGN luminosities are listed in Table 2.

\subsection{$R$--$L$ Relation}

The relation between the BLR size and quasar luminosity provides the basis for 
determining SMBH masses in high-redshift quasars/AGN with single-epoch
spectroscopy (for a recent review, see \citet{shen13}).
Accurate measurements of SMBH masses are particularly important in the
context of the SMBH and host galaxy co-evolution \citep{kor13}. The masses 
from RM can be calibrated from the local $M_{\rm BH}-\sigma_{\ast}$ relation
\citep[e.g.][]{ho15}. Our sample is still small, and would not improve  
the $R$--$L$ relation. On the other hand, the quasars in this sample are at 
relatively high redshifts, and thus may test the current
$R$--$L$ relation at $0.2<z<0.4$.

\begin{figure} 
\epsscale{1.1}
\plotone{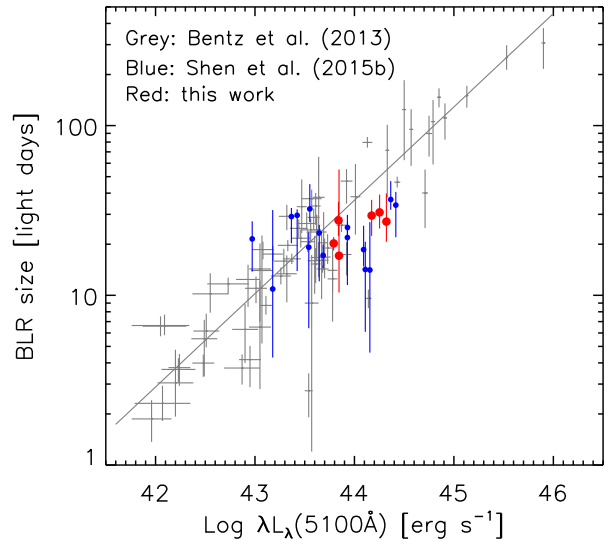}
\caption{The BLR size-luminosity relation in rest frame.
The gray symbols represent the
current $R$--$L$ relation from successful RM campaigns compiled by
\citet{ben13}. The blue circles represent recent results from SDSS-RM
\citep{shen15b}. Our results are shown as the red circles, and are roughly
consistent with the previous results.}
\end{figure}

Figure 9 shows the current $R$--$L$ relation from successful RM campaigns 
compiled by \citet{ben13}. Different emission lines have different ionization 
potentials, so the corresponding BLR sizes are different. The relation shown 
in Figure 9 was mostly built from \hb\ measurements of local AGN. We also plot 
the recent results on \hb\ and \mgii\ lags at $z\ge0.3$ (the blue 
circles) from the SDSS-RM project \citep{shen15b}.
Our results for the 6 quasars with lag detections are shown as the red circles.
They are roughly consistent with the $R$--$L$ relation derived from local AGN.
Our sample is primarily based on the \ha\ line, which has a shallower
ionization potential compared to \hb, and is thus expected to have a larger
BLR size. However, the difference between the \ha\ lag and \hb\ lag is
unclear, and may depend on quasar luminosity.
Several previous studies show that the difference ranges between 20\% and
50\% \citep[e.g.][]{kas00,kol03,ben10,haas11}. 
Given the scatter in the relation, our lag measurements are still consistent 
with the previous results.

We note that our 6 quasars occupy a small part of the parameter space in 
Figure 9. This is due to the small sample size and the strong target selection
bias. Our targets were selected to be bright, and they 
presumably have relatively large BLR sizes. If there were fainter quasars
in our sample, we would not be able to detect their time lags (and thus
they would not show up in Figure 9) because of the coarsely sample light
curves. On the other hand, much more luminous quasars do not show up in 
the figure either, since these quasars have much longer time lags and cannot
be detected in the IBRM duration, as we discussed in section 4.3.
Therefore, the consistency of our results with previous studies does not mean 
that we have validated the current $R$--$L$ relation at $0.2<z<0.4$.
A larger, unbiased sample covering a much larger parameter space is needed.

\section{SUMMARY AND FUTURE PROSPECTS}

We have presented our IBRM program, a photometric RM program with broad and 
intermediate-band photometry. The intermediate bands that we chose are 
centered at 8045, 8505, and 9171 \AA. They cover three wavelength ranges with 
relatively weak OH sky emission, thus imaging in these bands is very efficient.
Our sample consists of 13 quasars at redshift between 0.2 and 0.9. These
quasars were selected to have strong \ha\ or \hb\ emission lines that are 
located in one of the intermediate-bands. The IBRM campaign was carried out
with the 90Prime camera on the Bok telescope. The 90Prime has a large FoV and 
covered 13 quasars within five pointings/fields. The five fields were observed 
in the $i$, $z$, and intermediate-bands in 20--30 epochs. These epochs were
unevenly distributed in a duration of 5--6 months, so the cadence is not
optimal for RM experiments. By using a large number of standard stars for
each quasar, we achieved high accuracy on the photometric measurements.
The combination of the broad and intermediate-band photometry allows us to
precisely determine the light curves of the optical continuum and emission
line. We detected significant time lags between continuum and line emission
in 6 (out of 13) quasars in our sample. The time lags are consistent with
the $R$--$L$ relation derived from \hb\ in low-redshift AGN.

Photometric RM with intermediate-band photometry has two major advantages.
First, as with any implementation of photometric RM, it does not require 
spectroscopic observations, and can be easily performed with small telescopes.
Second, the bandwidth of an intermediate filter is narrow enough that
the line flux still contributes a significant fraction of the total flux
in the band. Meanwhile, it is wider than narrow bands so that it is 
possible to include more than one target (at similar redshifts) per
telescope pointing, which substantially increases observing efficiency.

Based on our experience from the IBRM program, we may increase our efficiency
and improve our success rate in future RM campaigns with intermediate-band 
photometry. We plan to carry out a larger RM program using the Near-Earth 
Object Survey Telescope (NEOST) in Xuyu, China. NEOST is a 1m telescope with 
a FoV of 9 deg$^2$. With such a large FoV, we can monitor several (up to
$\sim$10) quasars per telescope pointing. Our current IBRM experiment contains 
only 20--30 unevenly-distributed epochs. We will make more observations 
(40--50 epochs) with better cadence, which will largely increase the success 
rate and improve time lag measurements. Ideally, we can complete this RM 
campaign for 100 quasars with 60 nights (45 epochs in 6 months) on the NEOST 
telescope. 

We also plan to extend the baseline from 6 months to 18 months, with more
sparse sampling after 6 months. This is to explore higher-redshift and 
higher-luminosity quasars. The maximum redshift that the three intermediate 
bands can reach for \hb\ is roughly 0.9, which is much higher than the 
redshifts of the majority quasars shown in 
Figure 9. We will further extend this method other lines such as \mgii, 
although RM with \mgii\ is significantly more difficult because the line
is generally much weaker than \ha\ and \hb.

\acknowledgments

We acknowledge the support from a 985 project at Peking University, and the 
support from a Youth Qianren program through National Science Foundation of 
China. We would like to thank Y. AlSayyad, Y. Chen, K.D. Denney, S. 
Eftekharzadeh, Y. Gao, P.B. Hall, S. Jia, C.M. Peters, K. Ponder, J.A. 
Rogerson, R.N. Smith, and F. Wang for their help with the Bok observations.
The Pan-STARRS1 Surveys (PS1) have been made possible through contributions of the Institute for Astronomy, the University of Hawaii, the Pan-STARRS Project Office, the Max-Planck Society and its participating institutes, the Max Planck Institute for Astronomy, Heidelberg and the Max Planck Institute for Extraterrestrial Physics, Garching, The Johns Hopkins University, Durham University, the University of Edinburgh, Queen's University Belfast, the Harvard-Smithsonian Center for Astrophysics, the Las Cumbres Observatory Global Telescope Network Incorporated, the National Central University of Taiwan, the Space Telescope Science Institute, the National Aeronautics and Space Administration under Grant No. NNX08AR22G issued through the Planetary Science Division of the NASA Science Mission Directorate, the National Science Foundation under Grant No. AST-1238877, the University of Maryland, and Eotvos Lorand University (ELTE) and the Los Alamos National Laboratory.

{\it Facilities:}
\facility{Bok (90Prime)},
\facility{MMT (Red Channel spectrograph)},
\facility{Pan-STARRS1}

\end{document}